**The motility of normal and cancer cells in response to the combined influence of the**

**substrate rigidity and anisotropic microstructure**

Tzvetelina Tzvetkova-Chevolleau[1,2*], Angélique Stéphanou[2], David Fuard[1], Jacques Ohayon[2], Patrick Schiavone[1,3], Philippe Tracqui[2]

[1] LTM /CNRS/UMR5129, c/o CEA Grenoble, 17 rue des Martyrs, 38054 Grenoble cedex 9, France

[2] Laboratoire TIMC–IMAG, Equipe DynaCell, CNRS/UMR5525, Pavillon Taillefer, Faculté de Médecine de Grenoble - 38700 La Tronche, France

[3] Laboratoire TIMC–IMAG, Equipe GMCAO, CNRS/UMR5525, Pavillon Taillefer, Faculté de Médecine de Grenoble - 38700 La Tronche, France

*Corresponding author: Fax: +334 56 52 00 22

Tel: +334 56 52 00 43

E-mail address: tzvet@imag.fr

# Abstract

Cell adhesion and migration are strongly influenced by extracellular matrix (ECM) architecture and rigidity, but little is known about the concomitant influence of such environmental signals to cell responses, especially when considering cells of similar origin and morphology, but exhibiting a normal or cancerous phenotype. Using micropatterned polydimethylsiloxane substrates (PDMS) with tuneable stiffness (500kPa, 750kPa, 2000kPa) and topography (lines, pillars or unpatterned), we systematically analyse the differential response of normal (3T3) and cancer (SaI/N) fibroblastic cells. Our results demonstrate that both cells exhibit differential morphology and motility responses to changes in substrate rigidiy and microtopography. 3T3 polarization and spreading are influenced by substrate microtopography and rigidity. The cells exhibit a persistent type of migration, which depends on the substrate anisotropy. In contrast, the dynamic of SaI/N spreading is strongly modified by the substrate topography but not by substrate rigidity. SaI/N morphology and migration seem to escape from extracellular cues: the cells exhibit uncorrelated migration trajectories and a large dispersion of their migration speed, which increases with substrate rigidity.

## Introduction

The capacity of living cells to migrate in response to extracellular signals is crucial for many physiological processes such as embryonic development, wound healing, functionality of the immune [1, 2] and neural [3] systems. It also takes a crucial part in some pathological mechanisms such as tumour angiogenesis [4, 5]. While signals provided by the gradients of soluble chemoattractants are still considered as leading factors for orchestrating cell movements, additional guidance cues provided by physical and structural properties of extracellular matrices (ECM) have emerged over the last ten years as key parameters for guiding migration of cells. Particularly, matrix rigidity has been shown to strongly affect cell dynamics and physiological functions, including migration, division and apoptosis, differentiation, cytoskeleton organization, gene expression and phagocytosis [6-11]. Moreover, matrix stiffness has been shown to influence solid tumour formation and progression by an integrin-dependent regulation of the malignancy [5].

To understand the large variety of cell responses to the guidance signals provided by their surrounding substrate, many *in vitro* approaches using micro and nanotechnology-based tools have been developed [12]. Thus, the behaviour of different normal cells has been examined using a large variety of micropatterned substrates such as columns [13, 14], dots [15], pits [14, 16], pores [3], gratings [17] and random surface roughness [18], created by a variety of microlithography and microfabrication techniques. Thus, the use of surfaces with patterned topologies and adhesivity to organise cells morphologies has become a common strategy in tissue engineering. In contrast, methods using surface topography in addition to the stiffness of the culture substrate for organizing cell spatial distribution are quite recent [19] and remain less firmly established. Indeed, it remains largely unclear how cells sense combinations of different types of structural signals such as substrate rigidity and topography. Second, we do not know which of these two factors may predominantly drive the cell responses. Moreover, it is still unknown if this predominance may change according to the considered cell type or if the associated transduction pathways which orchestrate the cellular response lead to major differences in the motility behaviour of cancer cells compared to normal cells.

In this study, we used PDMS substrate with variable stiffness and patterned with micrometer-scale regions to evaluate the differential behaviour of normal fibroblasts (3T3 line) and transformed fibroblasts (SaI/N) on such surfaces. Using time-lapse videomicroscopy, we systematically investigated and compared the influence of the synergetic effects of substrate rigidity and topography on the morphological changes and motile behaviour of these normal and cancer cells when submitted to the same environmental signals.

## Materials and methods

### Cell cultures

Mouse 3T3 fibroblasts were cultured in Dulbecco's Modified Eagle's Medium (DMEM) (Sigma Aldrich) with 4500 mg/l L-glucose, supplemented with 10% foetal calf serum, 200mM L-glutamine, 0.25% penicillin/streptomycin. SaI/N cells (ATCC CRL-2544) are highly malignant fibroblastic cells, derived from mouse methylcholanthrene-induced sarcoma I tumour, which grow as solid tumours when inoculated subcutaneously [20]. SaI/N cells were maintained on Iscove's Modifed Dulbecco's Medium (IMDM) supplemented with 10% foetal calf serum and 2 mM L-glutamine. All cell culture reagents and media were obtained from Invitrogen (Invitrogen Ltd, UK). Both cell types were cultured under humidified atmosphere with 5% $CO_2$ and at 37°C. 3T3 and SaI/N cells were seeded at a density of 6500 cells/cm$^2$ onto the PDMS substrates.

**Fabrication of microstructured PDMS substrates**

Two different topographic patterns, with holes and lines, were fabricated in Silicium wafers using conventional 193nm photolithographic techniques followed by an etching process. The silicon wafers were etched down to 800nm, silanized with tridecafluoro-trichlorosilane in vapour phase to facilitate the release of the elastomer from the wafers after curing. $30mg/cm^2$ of a viscous liquid solution of silicone, polydimethylsiloxane (PDMS) (Sylgard 184, Dow Corning, Midland, MI) and its curing agent, were spilled over the silicon mould, levelled for 30 min at  room temperature and cured at 100°C  until the proper material rigidity is obtained. The PDMS polymers are then peeled off from the mould.

As observed with scanning electron microscopy, the replicas contain micro pillars with diameter of 1000nm, height of 800nm and centre-to-centre distance of 1600nm (Fig. 1A and C). Line patterning has been designed with lines having an equal width and inter-linear space of 750nm (Fig. 1B and D). After release from the mould, the replicas were treated with oxygen plasma (100 sccm $O_2$, in a LAM 9400 SE reactor at 5mT for 10 min) in order to make the PDMS surface hydrophilic (water contact angle in the range of 20-30°). Since water uptake in PDMS is very limited (less than 0,01g/g [21]), the hydrophilic properties of the substrates have been maintained in water for about 18*hrs* before starting the experiment. Then the substrates were washed 3 times with absolute ethanol and dried. The dried substrates were washed 3 times in PBS buffer and coated with fibronectin, a natural component of the cell ECM, for one hour temperature incubation at room temperature with 3.5µg/cm² fibronectin (Sigma-Aldrich). The fibronectin solution was removed and the substrates were washed 3 times in PBS and 3 times in culture media.

**Determination of PDMS Young's modulus**

The Young's moduli of PDMS substrates with different curing agent/PDMS polymer concentration ratios were measured by stretching unpatterned test samples with a stretching device (EMKA, Technologies, Paris). Briefly, each sample was cast in the form of a strip with size 9 cm. Each strip was uniaxially deformed along its longest axis by using suspended masses. Tension was applied in order to produce a strain. Under these conditions, the stress-strain relationship is linear for the tested sample and its Young's modulus $E$ was calculated according to the equation: $E=(FL)/(S\delta L)$, where $S$ and $L$ are the original cross-sectional area and length of the strip respectively, and $\delta L$ is the change in length produced by application of the tension $F$. At least three samples for each type of substrate were tested. The associated mean value of the Young's moduli are given with the standard deviation of the mean. Unpatterned PDMS substrates with Young's modulus of 500kPa ± 35kPa were obtained with a 0.5:10 w/w mixture of curing agent/PDMS polymer baked for 180min. PDMS substrate with Young's modulus of 750kPa ± 25kPa and 2000kPa± 100kPa were obtained using a ratio of 1:10 w/w mixture of curing agent/PDMS polymer further baked at 100°C for 15min or 180min, respectively [22].

For microstructured PDMS, we defined an homogenised Young's moduli of the substrate by following a homogeneization theory approach considering the volume fraction $V_f$ of PDMS within the sample. For an unpatterned PDMS substrate with Young's modulus $E_{PDMS}$, we thus defined in each case a homogenised Young's modulus of the patterned substrate as $E=V_fE_{PDMS}$. This provides a realistic first approximation of the rigidity experienced by the cultured cells crawling over pillars or lines where bending effects are limited due to width/height ratios close to 1. For micro pillars with diameter of 1μm, height of 0.8μm and centre-to-centre distance of 1.6μm, the volume fraction is $V_f \approx 0.31$, while for lines with equal width and inter-linear space, one gets $V_f=0.5$. Accordingly, for unpatterned PDMS with Young's moduli of 500kPa, 750kPa and 2000kPa, one will get pillar-patterned substrates with homogenised Young's moduli of 155kPa, 232kPa and 620kPa, respectively, and line-patterned substrates with homogenised Young's moduli of 250kPa, 375kPa and 1000kPa, respectively.

**Scanning Electron Microscopy**

SEM observations (Hitachi SEM 4000) were used to reveal the accuracy of the substrates microtopographies. Prior to SEM observation, the surfaces of the substrates were sputter-coated with a 2nm ± 1nm gold-palladium layer to make them electronically conductive and to avoid electronic charging during SEM imaging.

**Time-lapse videomicroscopy**

Time-lapse phase-contrast microscopy was performed by using an inverted Axiovert 135 microscope (Carl Zeiss, Jena, Germany) equipped with an incubation chamber. Cells were thus maintained during the time-lapse acquisition at a constant temperature of 37°C and constant 5% $CO_2$ in a wet atmosphere, in DMEM medium for 3T3 fibroblasts and IMDM medium for SaI/N fibrosarcomas. Images (Fig. 2) were taken with a CDD200 Cool Snap camera (Roper Scientific) and a 20X plan neofluar objective (Carl Zeiss) was used to collect images from different areas of the PDMS substrates.

Kinetics of initial cell spreading and polarisation were recorded by imaging every two hours six different regions, with image acquisition starting from one hour to nine hours after cell seeding. Cell tracking was started 10*hrs* after cell seeding and images were acquired every 10 minutes during 4 hours.

Image analysis was performed with Metavue software (Meta Imaging Series 6.1, Universal Imaging, Downingtown) and ImageJ software (National Institutes of Health, USA).

**Quantification of cell area and morphology**

Cells morphological changes were quantified by considering a shape factor given by the ratio between the approximated short (S) and long (L) cell axes. Cells with a S/L ratio lower than 0.75 were considered as polarised, and the time needed to reach this polarisation threshold is taken as a characteristic time of the cell polarisation. At least 150 cells were considered for each substrate.

Moreover, the value of the S/L ratio reached 9*hrs* after cell seeding, denoted as $S_9/L_9$, has been estimated for at least 20 cells per sample, together with the associated cell surface.

**Quantitative analysis of individual cell motility**

Motion of individual cells was analyzed from time-lapse recordings. The ImageJ software was used to determine the centroid position *{x(t_n),y(t_n)}* of each cell nucleus at each time point $t_n$, thus generating the cell migration path from the recorded *N* successive positions of the cell. The instantaneous cell speed *(v_n)* and cell path orientation *(α_n* ) were first quantified from cell migration paths by evaluating the following expressions:

$$v_n = \frac{\sqrt{d.d(t_n, \Delta t)}}{\Delta t}, \qquad \alpha_n = ar cos\left(\frac{x(t_n + \Delta t) - x(t_n)}{\sqrt{d.d(t_n, \Delta t)}}\right) \quad (1)$$

Moreover, the effectiveness *DE* of the cell motion has been introduced as the ratio between the distance *D*, separating the initial and final positions of the cell, and the total length *L* of the cell trajectory [23]. For each cell, the value *DE=D/L* has thus being computed, with:

$$D = \sqrt{d.d(t_0, N\Delta t)}, \qquad L = \sum_{n=0}^{N-1} \sqrt{d.d(t_n, \Delta t)}$$

(2)

The cell motion will be considered as more effective when the cell path almost follows a straight line, which corresponds to a value *DE* close to 1. On the contrary, random cell displacements would lead to smaller values of *DE* (*0≤DE≤1*). At least 25 different cells trajectories per sample were considered in this analysis.

In addition, a more refined analysis of cell motility has been performed by considering that individual cell trajectories may follow a persistent (or correlated) random walk (PRW). This hypothesis, introduced as a generalization of Brownian motion, is indeed often used to describe individual cell motion occurring at well-defined finite speed when persistence or inertia effects are not negligible [24].

Briefly, the persistent random walk (PRW) approach provides a unified way of characterizing cell motility by a two-parameter model which explicits the correlation between the directions of motion taken by the cell within successive time intervals. Practically, the squared displacement *d.d(t$_n$, i$\Delta t$)* of each cell is first computed over a time period *i. $\Delta t$* from the distance between the successive positions *{x(t$_n$), y(t$_n$)}* and *{x(t$_{n+i\Delta t}$),y(t$_{n+i\Delta t}$)}* of the cell, i.e.:

$$d.d(t_n, i\Delta t) = \left[ (x(t_n + i\Delta t) - x(t_n))^2 + (y(t_n + i\Delta t) - y(t_n))^2 \right]$$

(3)

Then, the mean-squared displacement $<d.d(t_k)>$ at a given time $t_k = k.\Delta t$ is obtained by averaging all the distances which can be computed when considering overlapping intervals of width $k$ covering all the cell positions, from the initial time $t_0$ up to the final observation time $t_{end} = N.\Delta t$ [24]. Application of this iterative procedure leads to the following expression of the mean-squared displacement:

$$< d.d(t_k) >= \frac{1}{(N - k + 1)} \sum_{i=0}^{N-k} d.d(t_i, k\Delta t) \tag{4}$$

Different theoretical approaches have shown that the mean squared displacement can be characterized by only two parameters, the mean speed $V$ and the persistence time $P$ according to the analytical model [24].

$$< d.d(t) >= 2V^2 P \left[ t - P(1 - e^{-t/P}) \right] \tag{5}$$

where the persistence time $P$ is a measure of the average time during which the cell maintains a given direction. However, this result has been established when considering an isotropic environment, and its relevance will be tested in our case when considering microstructured pillar substrates, for which anisotropy remains limited.

Thus, in the case of unpatterned and pillar-microstructured susbstrates, a non-linear least square fitting procedure has been used to identify from equations (4) and (5) the parameter set $(Vi, Pi)$ which provides the best fit to each cell trajectory (Fig. 3). In addition, the cell population fraction that fails to be described by the PRW model, has been quantified. In order to avoid larger values of $k$ (i.e. small number of intervals) leading to a biased averaging value in equation (2), only values of $k<2N/3$ are taken for the fitting procedure

**Statistics**

All results are reported as mean ± standard deviations of the mean. Analysis of the variances was performed using a two-way ANOVA test for independent samples developed by Vassar Colleges, USA, ©Richard Lowry 2001- 2007 (http://faculty.vassar.edu/lowry/anova2u.html). Statistical significance was set at $P < 0.05$.

# Results

The response of the considered two mouse cell lines with fibroblastic phenotype to combined influence of substrate topography and rigidity has been analysed with respect to the dynamics of cell spreading and morphological changes as well as the cell motility behaviour, including migration speed, orientation and effectiveness. For the sake of clarity, we will consistently use the terms soft, rigid, and very rigid for the designation of each type of substrates, patterned or unpatterned, made from PDMS materials with 500kPa, 750kPa and 2000kPa Young's modulus respectively.

**Morphological responses**

For each cell line, the progressive elongation (polarization) of the cells during and after spreading has been characterized by the fraction $f_{ss}$ of cells within the population that reaches a steady-state level of polarization lower than the shape factor criterion retained to defined a polarised shape ($S/L \leq 0.75$) and the time $t_{ss}$ taken by these cells to reach this steady-sate level.

For clarity, we do not report the overall set of curves showing the progressive increase of the fraction of cells becoming polarised with time. Instead, we summarize in Table 1 the values of the two parameters $f_{ss}$ and $t_{ss}$ for all the combinations of surfaces and rigidity we considered.

*Polarization kinetics*

In the case of 3T3 cells, the results presented in Table 1 show that the time of polarisation $t_{ss}$ significantly depends on the substrate properties. Whatever the substrate topography, the polarisation is faster on softer substrates ($t_{ss}$=3hrs), and increases with the substrate rigidity ($t_{ss}$=7-9hrs). In all cases however, a large fraction of the 3T3 cells (80%-90%) reaches the defined polarisation threshold.

In contrast, only a limited fraction of the cancer SaI/N cell population reaches a shape factor lower than the polarisation threshold, since about half of the cell population satisfies this condition (Table 1). Moreover, while the fraction of polarised 3T3 cells was rather insensitive to the substrate rigidity, the number of polarised SaI/N cells increases significantly with substrate rigidity and for all topographies: values of $f_{ss}$ are going from 40% to 55% for pillar and line microstructures, which is quite similar to the range of values (from 37% to 55 %) obtained with unpatterned substrates (Table 1).

Interestingly, the time $t_{ss}$ needed by SaI/N cells to reach these $f_{ss}$ values is strongly modified by the substrate topography, but unaffected by the substrate rigidity. The reverse property has been observed with normal 3T3 cells, the values of $t_{ss}$ being affected by the substrate rigidity and not by the substrate topography (Table 1). In the case of SaI/N cells, and by comparison with the unpatterned substrate case, it is particularly noticeable that the line microstructures tend to increase the kinetic of polarisation ($t_{ss}$ = 3hrs), while the pillar microstructures tend to lower it ($t_{ss}$=7hrs).

*Cell morphologies*

In order to correlate the above results with the projected surface of the cells on each substrate, we compiled in Figure 4 the mean cell surfaces and shape factors measured nine hours after cell seeding, when the cell population has reached the polarized steady-state level. The polarisation process of the 3T3 cells, as measured by the $S_9/L_9$ ratio is not significantly influenced by the modification in the substrate rigidity but is significantly influenced by the topographic changes in the substrates ($p < 0.005$ for all topographic comparisons). Otherwise 3T3 cell surface is sensitive to rigidity changes ($p < 0.001$), going from $700 \mu m^2$ on soft substrates down to $100 \mu m^2$ on very rigid substrate, No such effect exists with SaI/N cells, those final polarization and surfaces of SaI/N cells are both poorly affected by the substrate topography ($p > 0.05$) and rigidity ($p > 0.05$).

Figure 4 also shows that data for 3T3 cells all belong to the left part of the graph, which means that 3T3 cells are far more elongated than SaI/N cells for all considered substrates: the $S_9/L_9$ ratio is close to 0.1 and even reaches a lower value of 0.077 on very rigid substrates with line topography (Fig. 4). Unpatterned and pillar substrates give rise to a less pronounced cell response with a twice higher $S_9/L_9$ ratio of 0.15.

**Motility responses**

*Migration speeds*

To examine the influence of both substrate rigidity and topography on cells motile behaviour, we first computed using equation (1) the histogram of cell speed distribution within each type of cells and for the observation period of four hours. Figure 5 presents the set of histograms we obtained for the overall combinations of rigidities and topographies we considered. The range of possible cell speed values between 0 and 2 μm/min was divided into successive interval $I_p$ of 0.2 μm/min width. For each interval $I_p$, the corresponding bar height indicates the percentage of cells within the population that have migrated with a speed falling within the rage of values defined by $I_p$. Thus, a percentage value of 100% means that all the cells have reached the considered range of speed values at least once during their migration.

The left column in Figure 5 shows the distributions of migration speed computed for 3T3 cells. For unpatterned as well as for patterned surface, 80% of 3T3 cells exhibits a maximum migration speed lower than 0.6μm/min. Only a small fraction of the remaining cells (less then 20%) develops a maximum migration speed larger than1.4μm/min.

In contrast, SaI/N cells develop up to three times higher speeds (around 1.2 μm/min) than those observed for normal cells and independently on the topography. Only 35% of the cells develop small speeds in the range of 2-5 μm/min.

Moreover, figure 5 (right column) shows that the histograms computed for SaI/N tend to flatten out, which indicates a larger dispersion, and thus a weakest regulation, of the cell migration speed.

Interestingly, this dispersion is almost twice higher on very rigid substrates (Fig. 5, dark gray bars) when compared to the soft rigid substrates (Fig. 5, dark gray bars) for all types of substrate patterning. This indicates a clear influence of the substrate rigidity on the regulation of the cell migratory behaviour, since this modification of the histogram amplitude cannot be attributed to an intrinsic dispersion of the cell motility characteristics. A closer examination of the histograms of SaI/N cells also shows that the highest dispersion of cell speed is obtained with pillar microstructured substrates (middle figure, right column, Fig. 5), while lines topographies tend to channel the dispersion toward lowest migration speed (upper figure, right column, Fig. 5). Such influences have not been observed with normal 3T3 fibroblastic cells.

*Displacement effectiveness*

Evaluation of the cell displacement effectiveness from equations (2) reveals that this parameter is not significantly sensitive to the substrate topography, nor to the substrate rigidity. Figure 6 presents the histogram established for soft substrates (Fig.6), but similar histograms were obtained for rigid and very rigid substrates (data not shown). Let us remark that the effectiveness criterion does not automatically reflect the substrate anisotropy. Thus, line microstructures substrates will not systematically lead to the largest effectiveness value. Indeed, this value will decrease despite strong anisotropy if cells are moving back and forth along the lines. Considering again Figure 6, one can then notice that unpatterned or pillar substrates (which exhibit none or slight anisotropy or no anisotropy, respectively) can sustain migration paths as effective as line microstructured substrates. This remark still holds for SaI/N cells, but the displacement effectiveness of 3T3 cells is twice more efficient than for SaI/N cells ranging from 50% (unpatterned) to 60% (pillar), while SaI/N cells effectiveness reaches at 25% on soft substrate (Fig. 6).

*PRW migration*

The previous analysis does not take into account the potential correlation existing between the successive speeds and direction taken by the considered cell. We thus refine our quantitative characterization of the motile behaviour of the two fibroblastic cell types in the framework of persistent (or correlated) random walk motions, as defined in the Materials and Methods section.

First, we found that the PRW model can successfully describe the migration of both 3T3 cells and SaI/N cells, not only on unpatterned substrates, but also on pillars microstructured substrates, *i.e.* on weakly anisotropic surfaces. More precisely, we found a percentage of cells following a PRW-motion which is close to 90% for 3T3 cells, whatever the rigidity of the substrate (data not shown). A large fraction of 3T3 cells (~ 80%) have a speed *V* within the range of [0.1-0.5] μm/min. The PWR model can still be considered as relevant for describing SaI/N cells migration, but now only about 55% of these cancer cells follow a PRW motion, with a large dispersion of the values of *V* which is similar to the one reported in Figure 5.

Taking benefit of the two-parameter description of the overall cell trajectory provided by the PRW model (Equ. (5)), we further examine the persistence of the cell motion, *i.e.* the tendency for a cell to keep the same orientation of motion. We thus consider as a threshold value for the persistence time *P* the value *P\*=80min*, which corresponds to 1/3 of the observation period. Accordingly, a cell path for which the fitting procedure gives rise to an identified value of *P* larger than *P\** will be qualified as strongly persistent. Figure 7 presents the compiled results we obtained from the analysis of the time-lapse sequences of 3T3 and SaI/N cells migrating on unpatterned and pillar microstructured substrates using this persistence threshold. Interestingly, it appears that the percentage of 3T3 cells with strongly persistent motion depends on both substrate rigidity and topography (Fig.7). With unpatterned substrate increasing substrate rigidity decreases the percentage of cells with strongly persistent motion, from roughly 32% to 12%. In contrast, almost the same percentage of 3T3 (~40%) cells keeps a strongly persistent motion when cell migration takes place on pillar microstructured substrates. On the contrary, the percentage of SaI/N cells with strongly persistent motion remains always below 10%, whatever the substrate topography and rigidity (Fig. 7).

*Displacement orientation*

To gain further insights into the direction of migration taken by the cells relative to the substrate anisotropy, we use equation (1) to compute the direction of motion taken by each cell between two consecutive sampled positions. Results have been compiled in Figure 8, where the substrate directions given by the lines correspond to a direction angle of 90° (left column), while the pillars have been aligned perpendicularly (0° and 90° directions).

As expected, no preferential direction appears to be followed by the cells on unpatterned substrates. In contrast, the line topography clearly induces a significant oriented cell response, with cell migration in the direction of the lines. The sensitivity of this environmental cue is higher for 3T3 cells. Similarly, 3T3 cells also appear to orient at 90° and probably at 0° of the pillar position (Fig. 9). Thus, 3T3 cells migration over a network of pillars preferentially occurs along the horizontal and vertical directions given by the pillars spatial arrangement. Such an orientation is not observed for SaI/N cells.

The case of pillar substrates has to be analysed more carefully, since such substrate may appear as an intermediate topography between fully isotropic (unpatterned) environment and fully anisotropic line-type of substrate. In order to confirm the results obtained from figure 8 we compile for this particular substrates the so-called "wind roses" of cells trajectories, in which all cell trajectories start from the point (0,0). This macroscopic method confirmed the existence of two preferential migration directions, from the overall set of compiled trajectories of 3T3 cells (Fig 9). Those directions correspond to approximately 0 and 90° according to the pillar orientation. No such preferential directions of migration were observed for SaI/N cells cultured on pillar surface (data not shown).

## Discussion

Extensive studies have demonstrated that micropatterned surfaces provide physical cues that guide the migration of several cell types, including endothelial cells, fibroblasts or neurites [25, 26]. However, rather few studies are dealing with systematic analysis of the combined effects of topography and rigidity on cell behaviour [19, 27].

In this study, we used PDMS substrates with tuneable stiffness and specifically designed topographies to systematically investigate the concomitant effects of substrate topography and mechanical stiffness on the behaviour of normal fibroblasts (3T3) and transformed (SaI/N) fibroblasts isolated from fibrosarcoma. To our knowledge, this is the first time that cancer cell migration over microstructured substrate has been reported and compared to the behaviour of normal cells of same phenotype and species.

By applying the same environmental signals to these normal and cancer fibroblastic cells, we first established that the dynamic of cell spreading and progressive acquisition of an elongated morphology differ significantly between normal and cancer cells. Normal 3T3 cells reach more rapidly a polarised shape and keep a larger surface on soft substrates, whatever are the substrate topographies. In contrast, the polarisation kinetics and cell surface of transformed SaI/N fibroblastic cells appeared insensitive to variation in substrates stiffness. However, polarisation kinetics of SaI/N cells were affected by the substrate topography, those being faster on line substrates and delayed on pillar substrates when compared to unpatterned environment.

From extensive analysis of time-lapse videomicroscopy sequences, we then highlighted different motility responses of 3T3 and SaI/N to the physical cues provided by the different PDMS substrates. Particularly the percentage of 3T3 cells exhibiting a strongly persistent displacement on unpatterned substrates is almost divided by three when the PDMS rigidity becomes four times higher. The migration of 3T3 cells over micropillars substrate appears to be predominately governed by the directional cues given by the pillars alignments, this physical signals being sufficient to sustain the persistent motion of cells, even when the substrate rigidity is increased.

As a whole, the motile behaviour of SaI/N cells appears highly disorganized as demonstrated by the various cell parameters we analysed: the computed displacement effectiveness of the SaI/N cells is at least twice lower than the one computed for 3T3 cells and only roughly half of the SaI/N cells follow a correlated (persistent) random walk motion, either on unpatterned or pillar substrates.

Such differences may be explained by the different migration modes exhibited by normal 3T3 and cancer SaI/N cells as revealed by a close examination of their migration on micropatterned substrates. Increased spreading surface and polarisation of 3T3 cells suggest that these cells form and develop stable focal adhesions which favour the amoeboid type of migration that is clearly identified on time-lapse sequences (supplementary data 1). In agreement with the well admitted five-steps scenario of cell translocation [28], we can clearly observe the cell polarisation and cytoplasm membrane protrusion at the front, formation of strong adhesions with the substrate, cell body translocation and detachment of the rear part of the cell. The last step involves the recycling of adhesion receptors, which is not visible in our experimental conditions. This migrating mode is characterized by the development of important traction forces that can result in some cases in the sudden release of the cell adhesions that propels the cell forward. In contrast, (supplementary data 2), SaI/N cells, which keep a relatively round shape and limited spreading surface, seem to develop weaker types of adhesions with the PDMS substrates. This altered amoeboid migration mode, with reduced translocation phase, leads to higher cell speeds in this case.

A possible explanation for the oriented cell displacement of 3T3 cells could be proposed by considering that the substrate micropattern determines the spatial distribution of the fibronectin coating and its subsequent clustering with the cell integrin receptors. Such a spatial distribution of extracellular matrix (ECM) proteins has been recently shown to play an essential role in the control of cell spreading, migration dynamic and displacement orientation by influencing the formation of focal adhesions [29]. Reciprocally, focal adhesions are known to serve as membrane sensing entities [30] that control locally and globally adhesion-mediated cell signaling through Rho and Rac small GTPases [31], a family of proteins that are well known regulators of the actin cytoskeleton. Thus the geometry and patterns of the adhesive sites imposed by the substrate topography most probably drive directional cell migration by modulating Rho and Rac signalling pathways and thus, cell polarity, adhesion and traction forces [32]. In our experiments, we observe preferential cell displacements along horizontal and vertical directions of the pillars substrate. Since pillar spacing along the diagonals is roughly 1.4 times larger than along the horizontal and vertical axes, the favoured cell displacements at direction of 0° and 90° that we observed in our experiments could be linked to a denser spatial distribution of integrin ligands along those directions.

Closer comparison of our results in the light of reported data is limited by the variability of the experimental set-ups that have been used and by the associated huge heterogeneity that has been observed in the cell responses [33]. Nevertheless, some aspects deserve to be discussed. First, our results on the cell polarisation on lines and pillars are consistent with the results reported on microgroove topography [34] and on columnar microstructures fabricated by polymer demixing [35] or on three-dimensional sharp-tip microtopography [13]. In each case a clear relationship between the morphological parameters of normal cells and the substrate topography appears (e.g. lines topography constrains the cells to polarise and pillar topography favours cell spreading). Secondly, our results on cell motility are also consistent with the results obtained by Kaiser et al (2006) [34], who reported that 3T3 cells speeds are poorly affected by topographic changes of the extracellular environment. To our knowledge, no data is currently available on the speed of fibrosarcoma or other cancer cell types on micropatterned structures, which prevents further discussion.

**Conclusions**

This work provides for the first time a comparison of the synergetic influences of substrate rigidity and topography on normal and cancer cells behaviour. Our results highlight that a precise tuning between substrate micro-patterning and rigidity would allow for the control of cell morphodynamic parameters, such as cell surface, polarisation or speed. These can be readily incorporated into the design of microstructured substrates with tunable stiffness which can be utilized to direct cell response and associated mechanotransduction pathways. Specifically, the different morphological and dynamical behaviours of normal and cancer cells we observed in response to the rigidity and topography of the substrate may have strong insights in suggesting strategies for grading the metastatic phenotype of adherent cells, using calibrated micro-structured substrates. In addition, such microtechnologies may help the development and screening of new generation of drugs aiming at inhibiting cancer progression by discriminating cell sensitivity to the extracellular matrix topography and rigidity.

**Acknowledgments**

The mouse 3T3 fibroblasts cell line was kindly provided by Marc Block (Albert Bonniot Institute, Grenoble, France).

LEGENDS

Fig. 1.  Scanning electron microscopy images of micropatterned substrates with pillars (A) and lines (B). Contrast light microscopy images of 3T3 cells cultured on micropillars (C) and lines(D)  respectively.

Fig. 2.  Light microscopy images of the 3T3 fibroblast s(A,C,E)  and SaI/N fibrosarcoma (B,D,F) cells observed on the three different types of substrates: unpatterned,  pillars and lines respectively.

Fig. 3.  Illustration of the fitting procedure used to check the relevance of the PRW model of migration and to identify the two model parameters. The mean squared displacement <d.d> (solid points) of 3T3 cell cultured on 2000kPa PDMS substrate, was computed as a function of the number of intervals with increasing width (see equ. (2)). Solid lines correspond to the best least-squared fit to the computed mean squared displacements in the limit td < (2/3)$t_{end}$.

Fig. 4. Topography and rigidity effects on the cell surface and polarisation (presented by $S_9/L_9$ ratio) for 3T3-fibroblast and SaI/N-fibrosarcoma. The data present the results obtained from the topographied 500 and 2000kPa rigid substrates.

Fig. 5. Distribution of cells per speed intervals: 3T3 cells (left column), SaI/N cells (right column). Each bar represents the number of cells which have develop at least once the related cell speed during its migration. For each cell type, the distributions computed for two different substrate rigidities (soft and very rigid) have been compared.

Fig. 6. Mean cell displacement effectiveness on soft substrates with different topographies. 3T3-fibroblasts (in yellow) and SaI/N-fibrosarcoma (in gray).

Fig. 7. Percentage of cells showing persistence time P* larger than 80 min in the 3T3 and SaI/N populations (n> 30) on unpatterned and pillars topographies for soft and rigid PDMS substrates.

Fig. 8. Distributions of the angular directions of the 3T3 and SaI/N cells trajectories on unpatterned (A) and linear substrates (B).

Fig. 9. Compilation of 30 trajectories exhibited by 3T3 cells migrating on rigid pillar substrates. The dominant directions of migration followed by the cells are highlighted by dotted lines. The pillar network orientation is indicated by the arrows.

**Table 1**

Fraction of the 3T3 and SaI/N (with grey background) cells having reached a steady-state below the polarisation threshold $S/L \leq 0.75$ and corresponding time $t_{ss}$ to reach this steady-state. Values are percentage ± standard deviations of the mean.

| | Unpatterned topography | Pillar topography | Lines topography |
|---|---|---|---|
| **Soft** | 83%±4%(3h) | 90% ± 4%(3h) | 90% ± 1%(3h) |
| | 37%±1%(5h) | 40% ± 2%(7h) | 40% ± 1%(3h) |
| **Rigid** | 78%±3%(8h) | 84% ± 3%(7h) | 86%±6% (4h) |
| | 50% ± 2% (5h) | 47%±4%(7h) | 40% ± 1%(3h) |
| **Very rigid** | 83% ± 2%(9h) | 83%±7% (9h) | 85%±3% (7h) |
| | 55%±1% (5h) | 56% ± 4%(7h) | 54%±8% (3h) |

**Fig. 1.**

A

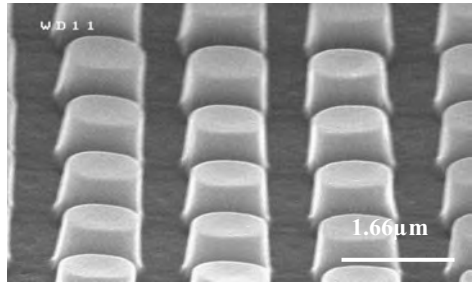

B

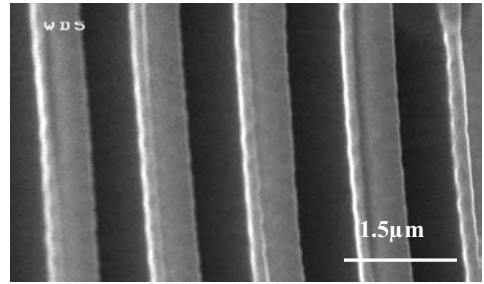

C

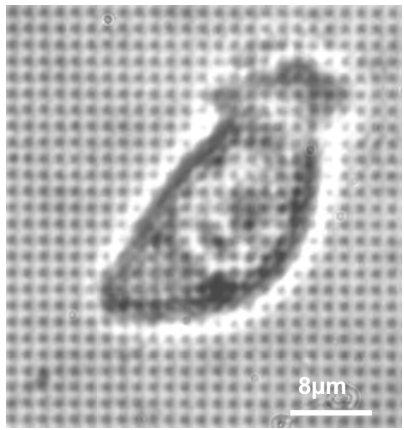

D

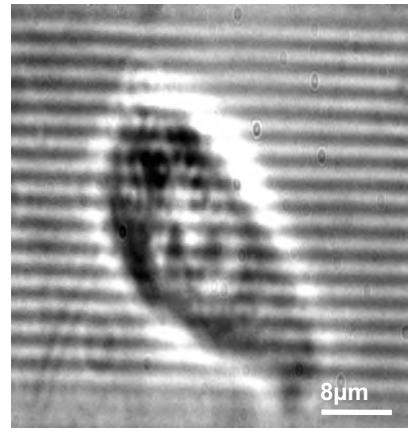

Fig. 2.

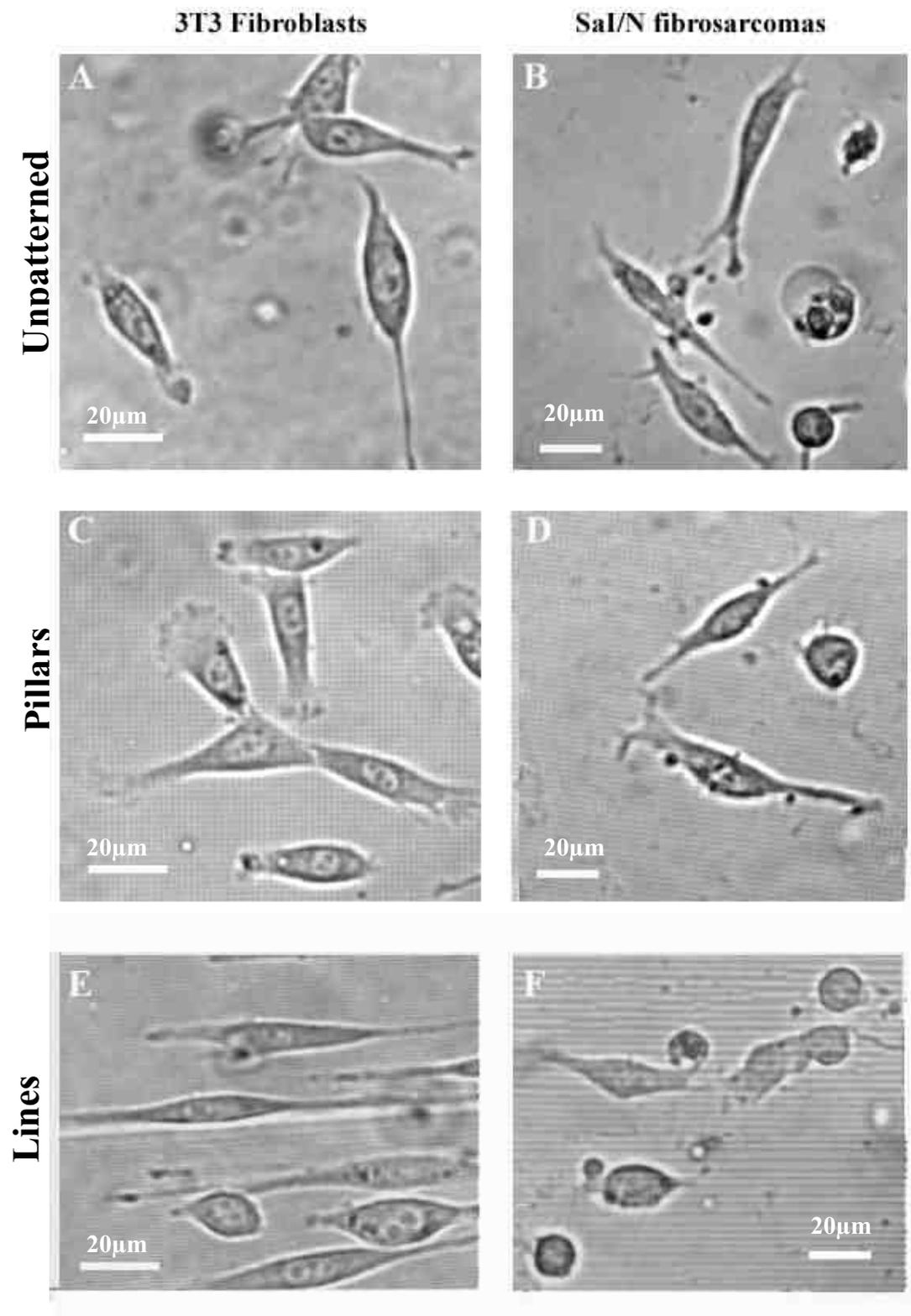

Fig. 3.

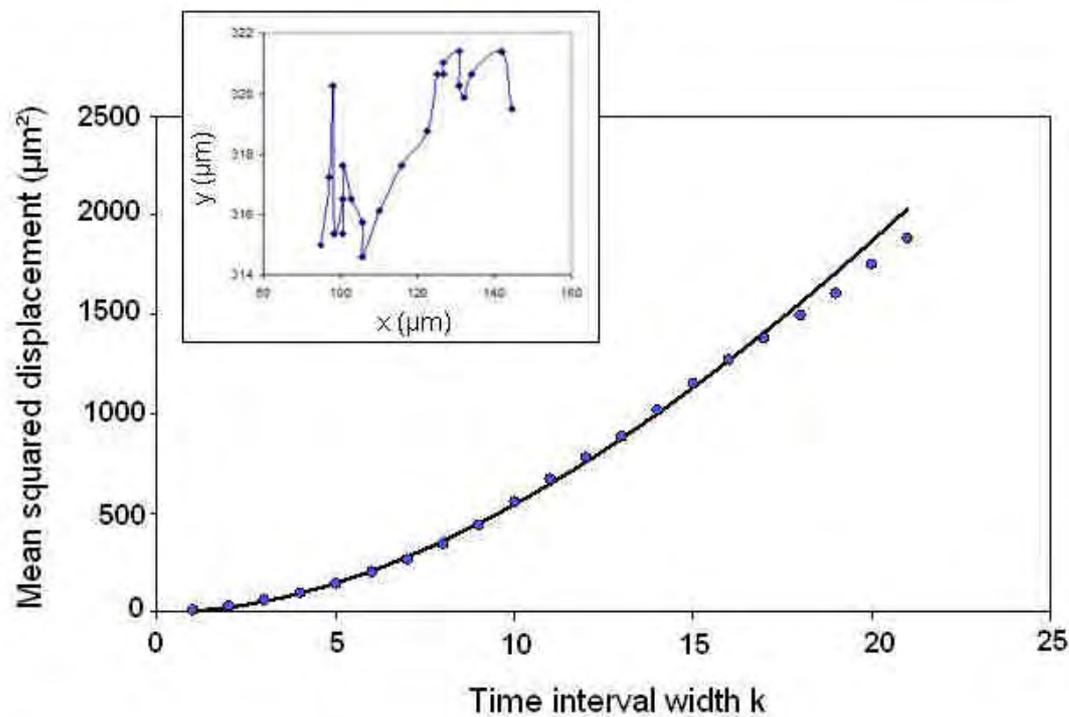

Fig. 4.

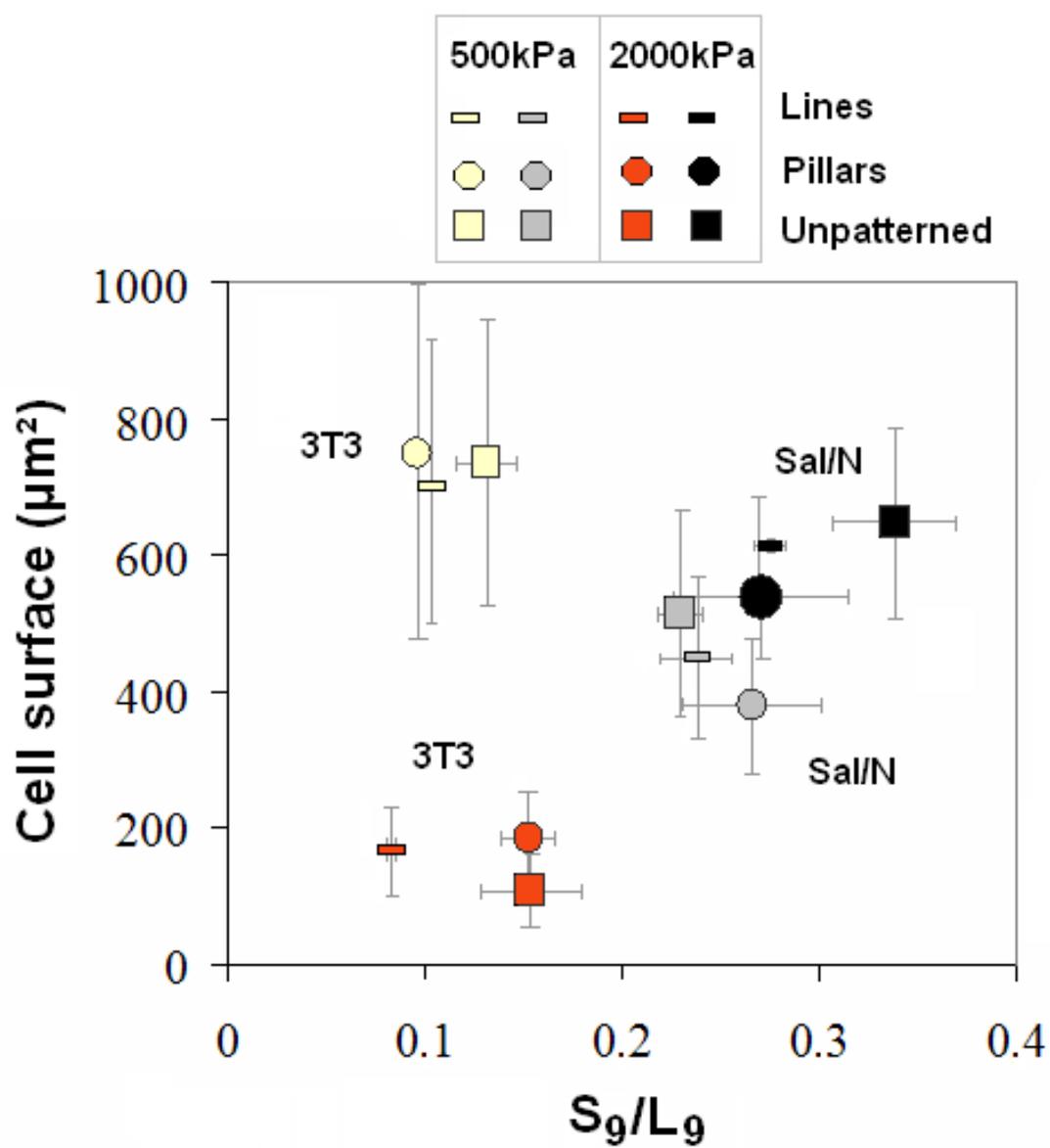

**Fig. 5.**

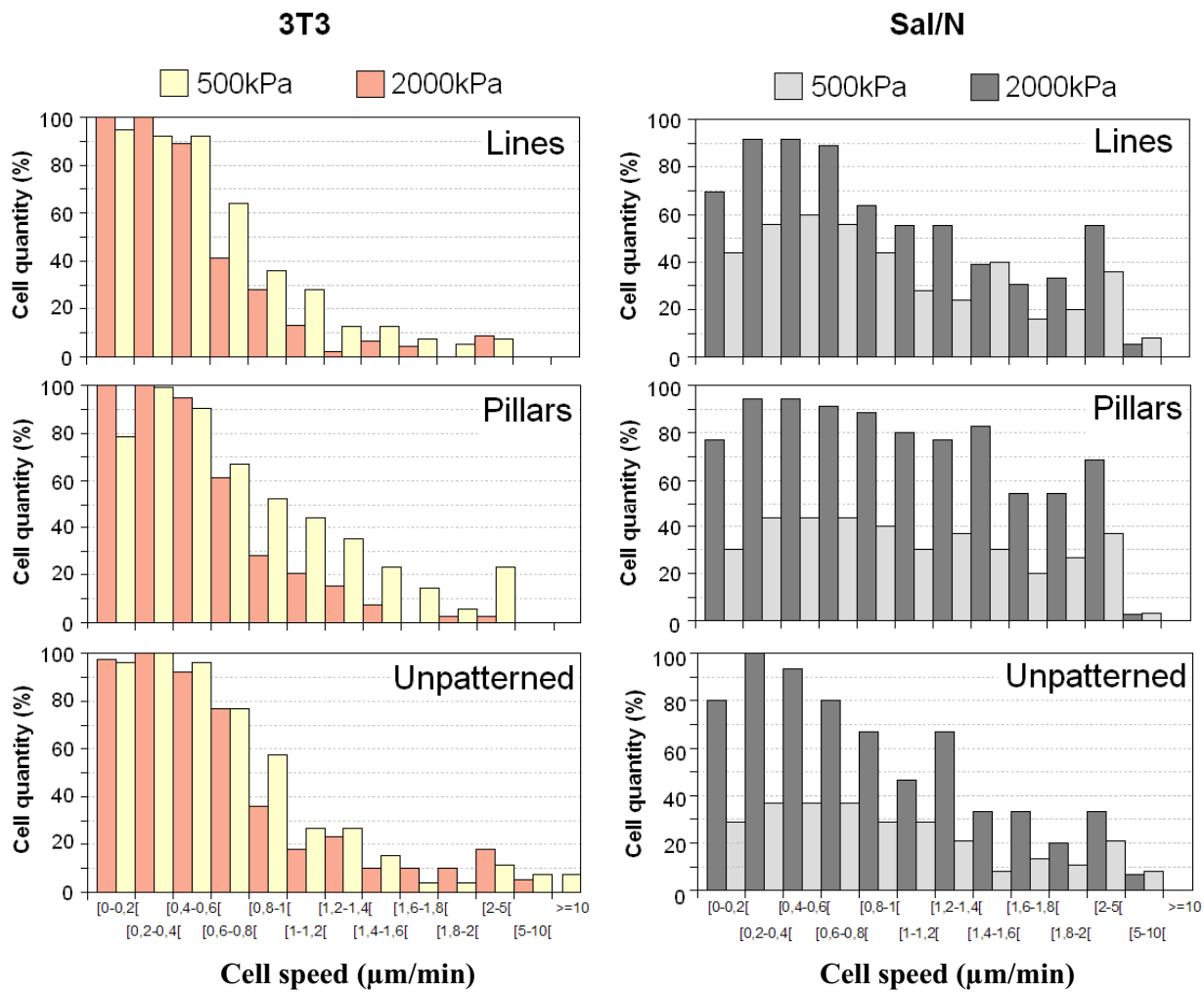

Fig. 6.

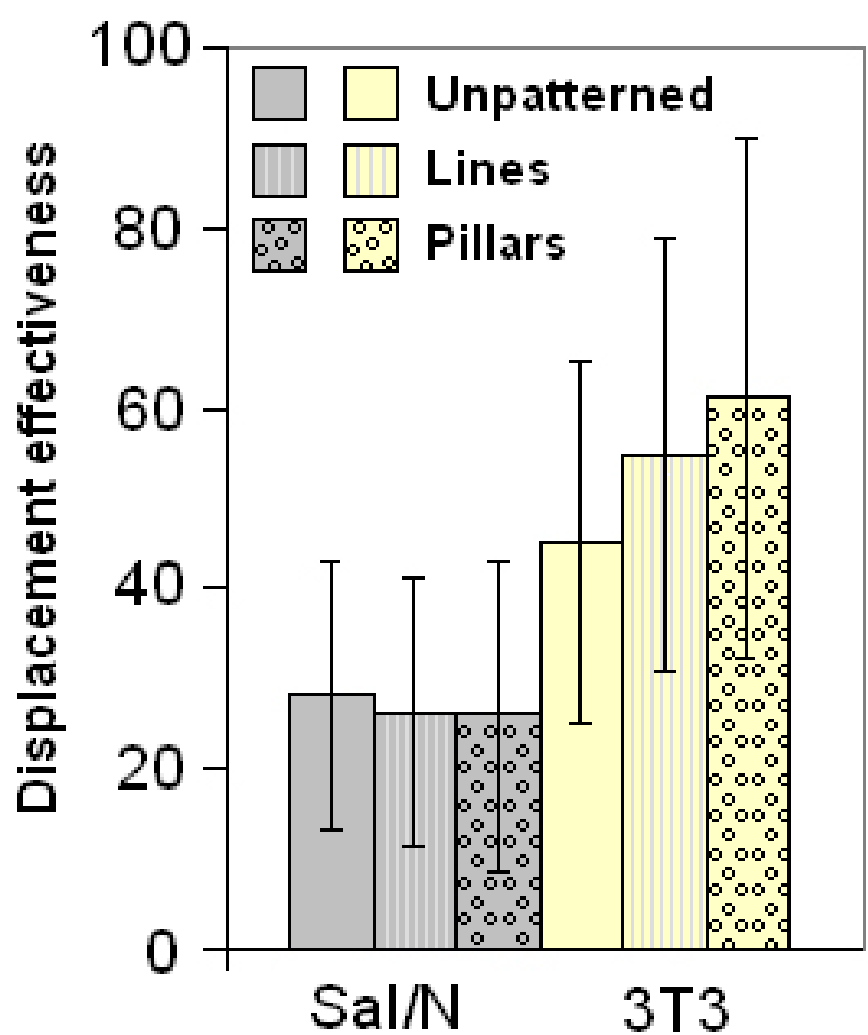

**Fig. 7.**

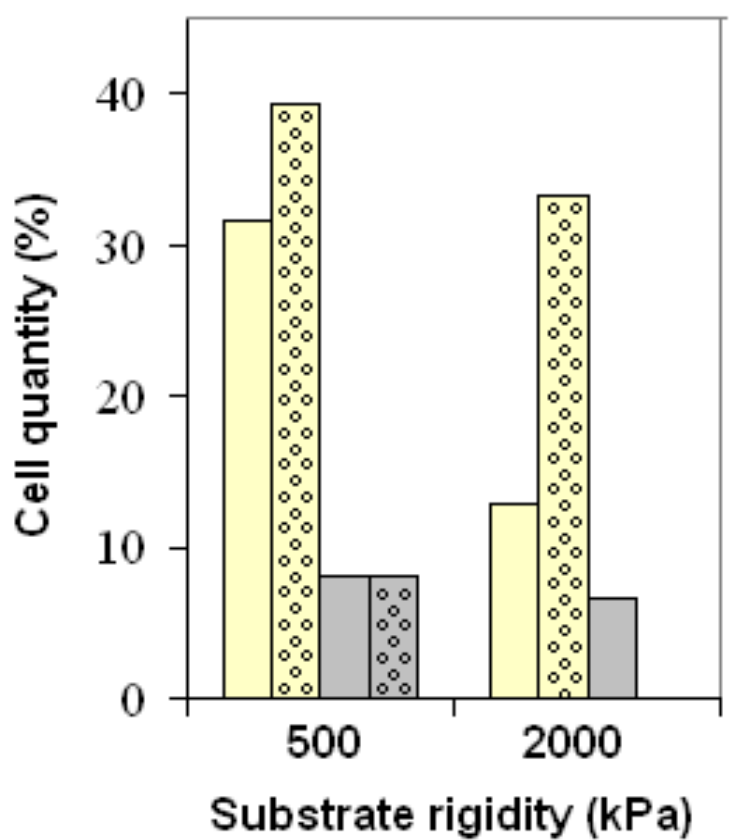

**Fig. 8.**

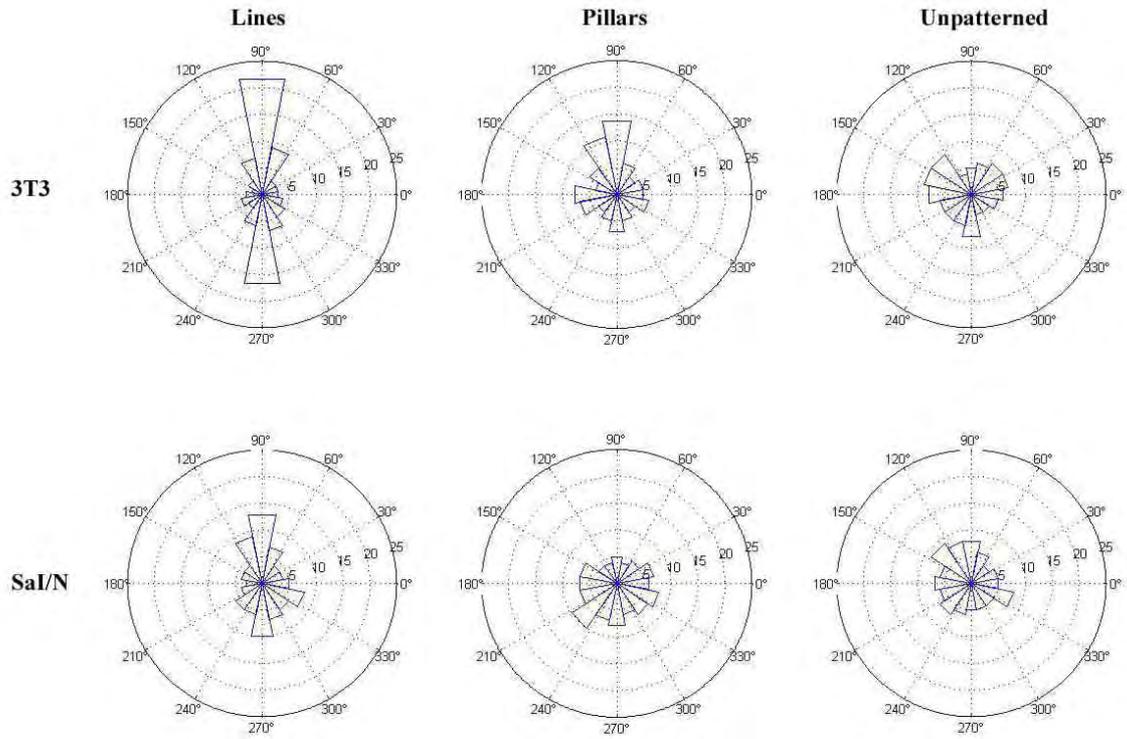

**Fig. 9.**

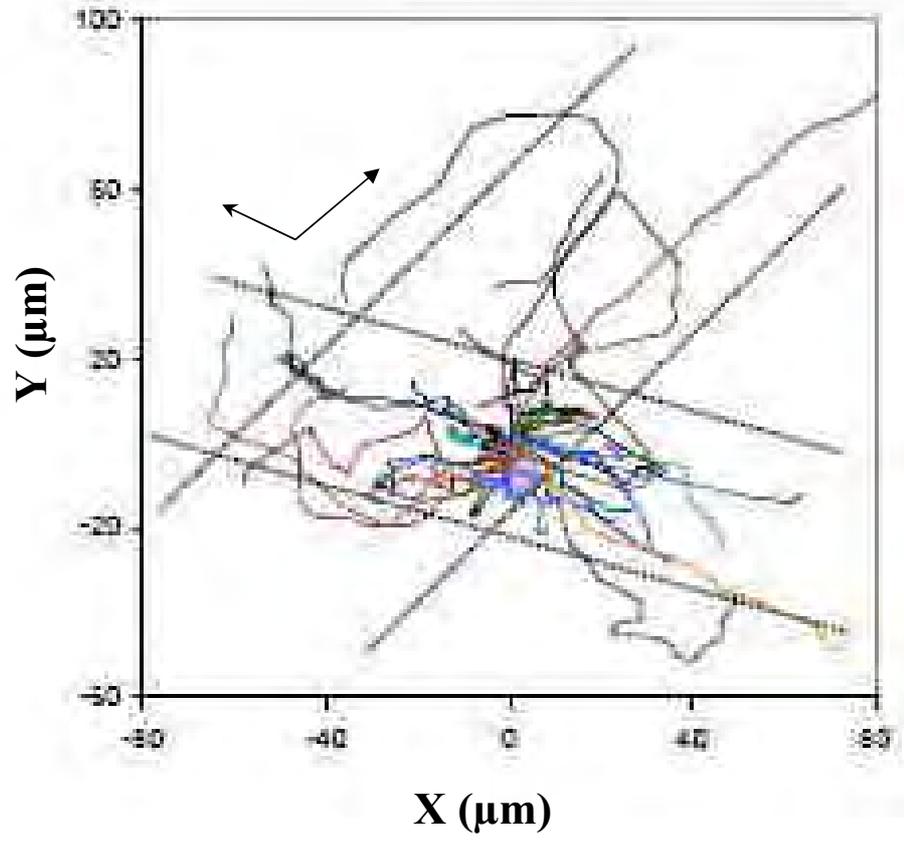